\documentclass[article,fleqn,showpacs,nofootinbib,preprint]{revtex4}
\usepackage{graphicx}
\usepackage{amssymb}
\usepackage{amsmath}
\usepackage{bm}
\usepackage{color}

\newcommand{\be}{\begin{eqnarray}}
\newcommand{\ee}{\end{eqnarray}}

\begin{document}

\begin{flushright}
{\tt arXiv:YYMM.NNNN}
\end{flushright}

\setcounter{page}{1}
\title{Four dimensional string solutions in Ho\v{r}ava-Lifshitz gravity}
\author{Inyong \surname{Cho}}
\email{iycho@snut.ac.kr}
\affiliation{School of Liberal Arts, Seoul
National University of Technology, Seoul 139-743, Korea}
\author{Gungwon \surname{Kang}}
\email{gwkang@kisti.re.kr} \affiliation{Korea Institute of Science
and Technology Information, 335 Gwahak-ro, Yuseong-gu, Daejeon
305-806, Korea}

\date[]{}

\begin{abstract}
We investigate string-like solutions in four dimensions based on
Ho\v{r}ava-Lifshitz gravity.
For a restricted class of solutions where the Cotton tensor
vanishes, we find that the string-like solutions in Einstein gravity
including the BTZ black strings are solutions in
Ho\v{r}ava-Lifshitz gravity as well.
The geometry is warped in the same way as in Einstein gravity,
but the ``conformal" lapse function is not constrained in
Ho\v{r}ava-Lifshitz gravity.
It turns out that if $\lambda \ne 1$, there exist no other solutions.
For the value of model parameter with which
Einstein gravity recovers in IR limit (i.e., $\lambda=1$),
there exists an additional solution of which the conformal
lapse function is determined.
Interestingly, this solution admits a
uniform BTZ black string along the string direction,
which is distinguished from the warped BTZ black string in Einstein gravity.
Therefore, it is a good candidate for the test of the theory.
\end{abstract}



\maketitle

\section{Introduction}
Very recently Ho\v{r}ava proposed a new quantum gravity theory which
is improved in renomalizability in
UV~\cite{Horava:2009uw,Horava:2008ih}. This theory treats space and
time on an unequal footing, and the theory becomes nonrelativistic.
This Ho\v{r}ava-Lifshitz theory attracted much attention in gravity
theory~\cite{Horava:2009if,Volovich:2009yh,Takahashi:2009wc,
Kluson:2009sm,Sotiriou:2009gy,Orlando:2009en,Charmousis:2009tc} and
particularly in cosmology~\cite{Calcagni:2009ar,Kiritsis:2009sh,
Brandenberger:2009yt,Piao:2009ax,
Gao:2009bx,Mukohyama:2009gg,Park:2009zr,Kim:2009dq,Wang:2009rw,
Wang:2009yz,Saridakis:2009bv,Bogdanos:2009uj}
and black-hole physics~\cite{Lu:2009em,Ghodsi:2009rv,Nastase:2009nk,
Cai:2009ar,Colgain:2009fe,Cai:2009pe,Myung:2009dc,
Myung:2009va,Lee:2009rm,Kehagias:2009is,Mann:2009yx,
Ghodsi:2009zi,AyonBeato:2009nh}.

In the Ho\v{r}ava-Lifshitz theory, the space and time are scaled
differently, \be {\rm x} \to b {\rm x}, \qquad t \to b^zt, \ee and
according to this rescaling, the space and the time dimensions are
\be [{\rm x}] = -1, \qquad [t] = -z \ee in mass dimension. The $z=3$
case corresponds to the three spatial dimensions, and is
power-counting UV renormalizable.

Using the ADM formalism, the four dimensional metric is written as
\be
ds_4^2= - N^2  dt^2 + g_{ij} (dx^i - N^i dt) (dx^j - N^j dt),
\ee
and the dimensions for each metric coefficients are
\be
[g_{ij}] = 0, \qquad [N_i] = z-1, \qquad [N] = 0.
\ee
In this ADM formalism, the Einstein-Hilbert action is given by
\be
S_\text{EH} =
\frac{1}{16\pi G} \int d^4x \sqrt{g} N (K_{ij} K^{ij} - K^2 + R
-2\Lambda),
\ee
where $G$ is Newton's constant, $R$ is the 3D Ricciscalar,
and $K_{ij}$ is the extrinsic curvature defined by
\be
K_{ij} = \frac{1}{2N} (\dot g_{ij} - \nabla_i N_j - \nabla_j N_i).
\ee

In the Ho\v{r}ava-Lifshitz gravity for $z=3$,
the kinetic part for the action is
\be
S_K=\frac{2}{\kappa^2}\int dt d^3x \sqrt{g} N
(K_{ij}K^{ij}-\lambda K^2),
\ee
where $\kappa$ and $\lambda$ are
dimensionless couplings, and for $\lambda =1$ the kinetic part
becomes that of the Einstein-Hilbert action. The remaining terms
correspond to the nonrelativistic potential term which satisfy the
so called ``detailed balance" condition,
\be
S_V=\frac{\kappa^2}{8}\int dt d^3x \sqrt{g}N {\cal E}^{ij}{\cal
G}_{ijkl} {\cal E}^{kl}, \qquad \sqrt{g}{\cal E}^{ij}\equiv \frac{\delta
W[g_{kl}]}{\delta g_{ij}},
\ee
where ${\cal G}_{ijkl}$ is the inverse of the De Witt metric,
\be
{\cal G}^{ijkl} = \frac{1}{2}
\left( g^{ik}g^{jl} + g^{il}g^{jk} \right) -\lambda g^{ij}g^{kl},
\ee
and $W[g_{ij}]$ is a three dimensional Euclidean action,
\be
W=\frac{1}{w^2}\int \omega_3(\Gamma)+\mu \int d^3 x
\sqrt{g}(R-2\Lambda_W),
\label{WWW}
\ee
where the first term is the
gravitational Chern-Simons term with the dimensionless coupling $w$,
and the second term is a three dimensional Einstein-Hilbert term
with a coupling $\mu$ of dimension 1 and a three dimensional
cosmological constant $\Lambda_W$ of dimension 2.

With the above kinetic and potential terms, the 6th-order action for
Ho\v{r}ava gravity becomes
\be
S&=&\int dtd^3{\rm x}\, ({\cal L}_0 + {\cal L}_1), \label{ActionHG}\\
{\cal L}_0 &=& \sqrt{g}N\left\{\frac{2}{\kappa^2}(K_{ij}K^{ij}
-\lambda
K^2)+\frac{\kappa^2\mu^2(\Lambda_W R -3\Lambda_W^2)}{8(1-3\lambda)}\right\}, \\
{\cal L}_1&=& \sqrt{g}N\left\{\frac{\kappa^2\mu^2
(1-4\lambda)}{32(1-3\lambda)}R^2 -\frac{\kappa^2}{2w^4} \left(C_{ij}
-\frac{\mu w^2}{2}R_{ij}\right) \left(C^{ij} -\frac{\mu
w^2}{2}R^{ij}\right)\right\},
\ee
where $C^{ij}$ is the Cotton tensor defined by
\be
C^{ij}=\epsilon^{ik\ell}\nabla_k\left(R^j{}_\ell
-\frac14R\delta_\ell^j\right)=\epsilon^{ik\ell}\nabla_k R^j{}_\ell
-\frac14\epsilon^{ikj}\partial_kR.
\ee
Comparing ${\cal L}_0$ with
that of general relativity in the ADM formalism, the speed of light,
the Newton's constant and the cosmological constant are related with the
model parameters as
\be
c=\frac{\kappa^2\mu}{4}
\sqrt{\frac{\Lambda_W}{1-3\lambda}}, \qquad
G=\frac{\kappa^2}{32\pi\,c},\qquad \Lambda=\frac{3}{2} \Lambda_W .
\label{c}
\ee
In order for the speed of light to be real,
$\Lambda_W <0$ when $\lambda >1/3$. Performing an analytic
continuation, $\mu \to i\mu$, $w^2 \to -iw^2$, we can have
$\Lambda_W >0$ when $\lambda >1/3$, which makes the action
consistent. After the analytic continuation, the potential terms
change their signature while the kinetic terms remain unchanged.

The field equations from the action~\eqref{ActionHG} were derived in
Refs.~\cite{Kiritsis:2009sh,Lu:2009em}. We summarize the results in
Appendix for further convenience.

In this work, we investigate an axially symmetric system in
Ho\v{r}ava-Lifshitz gravity. We solve field equations and obtain
static solutions with vanishing Cotton tensor.
We discuss that a BTZ-type black-string solution is possible for a
specific value of the model parameter in this theory.

\section{BTZ black strings in Einstein gravity with cosmological
constant}

In this section we briefly review the BTZ black-string solutions in
four dimensional Einstein gravity with a cosmological
constant $\Lambda_4$~\cite{Kang:2006zh}.
The Einstein equation in four dimensional spacetime in the presence
of the cosmological constant can be written as \be R_{MN} =
\Lambda_4 g_{MN}. \ee One can easily see that any metric satisfying
the three dimensional Einstein equation in the presence of the 3D
cosmological constant $\Lambda_3$, \be \tilde{R}_{\mu\nu} = 2
\Lambda_3 \gamma_{\mu\nu}, \label{3DEinstein} \ee can be embedded
into the 4D warped geometry given by \be ds^2 = W^{-2}(z) \left[
\gamma_{\mu\nu}(x^\sigma) dx^{\mu}dx^{\nu} +dz^2 \right]. \ee Here,
the warp factor satisfies \be \frac{\ddot{W}}{W}
-\left(\frac{\dot{W}}{W}\right)^2 -\frac{\Lambda_4}{3W^2} = 0,
\qquad {\rm where} \qquad \dot{} \equiv \frac{d}{dz}, \label{WarpGR}
\ee and $\Lambda_3$ is given by \be \Lambda_3 = \frac{\ddot{W}}{W}.
\label{3DC} \ee

If the 4D bulk cosmological constant is negative, $\Lambda_4<0$,
there are three types of solutions to Eq.~\eqref{WarpGR},
\begin{equation}
W(z)=
\left\{
  \begin{array}{ll}
    \sqrt{\frac{-\Lambda_4}{3\Lambda_3}} \sinh \sqrt{\Lambda_3}(z -z_0)
& \hbox{\qquad {\rm for} \quad $dS_3 \, (\Lambda_3 > 0)$,} \\
    \sqrt{\frac{-\Lambda_4}{3}} (z -z_0)
& \hbox{\qquad {\rm for} \quad $M_3 \, (\Lambda_3=0)$,} \label{warp2}\\
    \sqrt{\frac{\Lambda_4}{3\Lambda_3}} \sin \sqrt{-\Lambda_3} (z -z_0)
& \hbox{\qquad {\rm for} \quad $AdS_3 \, (\Lambda_3 < 0)$.}
  \end{array}
\right.
\end{equation}
where $z_0$ is an integral constant, and $\Lambda_3$ is related with
the other integration constant through Eq.~\eqref{3DC}.
On the other hand, if the bulk cosmological constant is positive,
$\Lambda_4 > 0$, the only possible solution is given by
\be
W(z)= \sqrt{\frac{\Lambda_4}{3\Lambda_3}} \cosh
\sqrt{\Lambda_3}(z -z_0),
\label{warp4}
\ee
where $\Lambda_3$ is positive only in this case.

In (1+2) dimensions, the static circularly-symmetric  solution
to the 3D Einstein equation~\eqref{3DEinstein} is given by
\be
ds_3^2 = -\left( -M -\Lambda_3 r^2 \right) dt^2 +\frac{dr^2}{-M
-\Lambda_3 r^2} +r^2d\theta^2 .
\label{BTZ}
\ee
For $\Lambda_3 <0$ with $M>0$, this is the so called BTZ black hole.
The corresponding 4D solution does not have a translation symmetry
along the extra $z$-direction.
Due to the warp factors in Eqs.~(\ref{warp2})-(\ref{warp4}),
for the case of $\Lambda_3<0$,
the warped geometry is the BTZ black string in 4D Einstein gravity,
\be
ds^2 = \frac{3\Lambda_3/\Lambda_4}{\sin^2 \sqrt{-\Lambda_3}(z-z_0)}
\left[ -\left( -M -\Lambda_3 r^2 \right) dt^2 +\frac{dr^2}{-M
-\Lambda_3 r^2} +r^2d\theta^2 +dz^2 \right] .
\label{BTZinGR}
\ee

\section{Gravity of String Model}
In this section, we consider a string-like object in
Ho\v{r}ava-Lifshitz gravity. We consider a static system with axial
symmetry in four dimensions. The metric ansatz is then
\be
ds^2 = W^{-2}(z)\left[ -\tilde{N}^2(r)dt^2 + \frac{dr^2}{f(r)}
+r^2d\theta^2 +dz^2\right],
\label{metric}
\ee
where $W(z)$ is the warp factor.
In the four dimensional Einstein gravity with vanishing cosmological
constant, it is known that there is no stationary black-string
solution. The topology theorem for stationary black holes in 4D
simply contradicts with
the existence of such a black-string configuration.
Presumably, this is also related with
the fact
that there is no black-hole solution in the 3D Einstein gravity so
that it is impossible to obtain a black-string configuration in 4D
by arranging 3D black holes along one spatial direction.
In the presence of a cosmological
constant, however, black-hole solutions exist in 3D, for example,
BTZ black holes. By foliating these 3D BTZ black holes along one
direction, one can obtain black-string solutions in four dimensions
as shown above. In this section, we search for such BTZ-type
black-string solutions in the 4D Ho\v{r}ava-Lifshitz gravity.

Since we are dealing with a static system without a momentum flow
($N^i=0$), the extrinsic curvature vanishes, $K_{ij}=0$. Therefore,
the field equations in Appendix are simplified, and the equation
from $\delta N^i$ is absent. Since the kinetic terms involved with the
extrinsic curvature are missing, the resulting field equations are
the same regardless of the ranges of $\Lambda_W$ and $\lambda$
(i.e., after the analytic continuation, the equations remain
unchanged).

Now, we solve field equations \eqref{eom1} and \eqref{eom3}.
Unfortunately, however, we are unable to solve those equations
fully analytically. Instead,
we consider the situation in which the Cotton tensor
vanishes.~\footnote{This case corresponds to the Ho\v{r}ava-Lifshitz gravity
having no contribution from the gravitational Chern-Simons term.}
Note that the Cotton tensor identically vanishes for the static
system with spherical symmetry studied in Ref.~\cite{Lu:2009em}.
For the system with axial symmetry with the metric
ansatz~\eqref{metric}, however, the Cotton tensor does not vanish
identically, and the nonvanishing components are
\be
C_{\theta z} = C_{z \theta} = -\frac{1}{4}W\sqrt{f} \left(f''
-\frac{f'}{r}\right),
 \ee
where the prime denotes the derivative with respect to $r$.
The solution to $C_{ij}=0$ is
\be
f(r) = -M -\alpha r^2, \label{A}
\ee
where $M$ and $\alpha$
are integration constants. Later, we shall see that $M$ is arbitrary
playing the role of the mass, and that
$\alpha$ plays like an effective cosmological constant of the
$(1+2)$ dimensional spacetime transverse to the $z$-direction.

By substituting the metric function $f(r)$
in Eqs.~\eqref{eom1} and \eqref{eom3} with Eq.~\eqref{A},
we obtain a simpler set of equations
shown in Eqs.~(\ref{Ett})-(\ref{Ethz}).
Although Eq.~(\ref{Ett}) provides a decoupled equation
involving the warp factor $W(z)$ only,
it is unlikely to be solved easily. Other equations consist of
both $\tilde{N}(r)$ and $W(z)$ functions and their derivatives, but
all of them are separable.
One can easily check that all solutions for
the warp factor in Einstein gravity given in Eqs.~(\ref{warp2})-(\ref{warp4})
with $\tilde{N}^2= -M -\Lambda_3 r^2$ and $\alpha = \Lambda_3$,
become the solutions in Ho\v{r}ava-Lifshitz gravity as well.
We shall show below that indeed
there exists only this class of solutions in Ho\v{r}ava-Lifshitz
gravity, but the conformal lapse function $\tilde{N}$ is
not necessarily constrained.
In addition, for the special case of $\lambda = 1$,
there exists a BTZ-type black-string solution, but interestingly a
constant warp factor is also allowed; the space without
being warped along the string direction also exists.

\subsection{Solution with $\tilde{N}(r)$ unconstrained}
Assuming that $\lambda \not= 1$, one can use Eq.~(\ref{Ett}) to
replace $\ddot{W}$ and its higher derivatives in other equations.
Then the subtraction between Eqs.~(\ref{Err}) and (\ref{Ethth})
presents our {\it master} equation
 \be
\left[ 1 +\lambda \left( -3 \pm \sqrt{2(3\lambda -1)}\right) \right]
\left[ {\dot{W}}^2 -\alpha W^2 +\Lambda_W \right] \left[ r \left( -M
-\alpha r^2 \right) \tilde{N}'' +M \tilde{N}' \right] = 0.
\label{Eththmrr}
 \ee
The first term in the above equation vanishes when $\lambda = 1/3,1/2$,
or $1$. Thus, if
$\lambda \not= 1/3, 1/2, 1$, we have either \be {\dot{W}}^2 -\alpha
W^2 +\Lambda_W = 0, \label{Weq} \ee or \be r \left( -M -\alpha r^2
\right) \tilde{N}'' +M \tilde{N}' = 0. \label{Neq} \ee Both
equations can be solved analytically.

The solution to Eq.~(\ref{Weq}) is given by
\be
W(z) = C_1 e^{\sqrt{\alpha} z} +C_2 e^{-\sqrt{\alpha} z} \qquad {\rm
with} \qquad C_1C_2 = \frac{\Lambda_W}{4\alpha}.
\label{WarpsolUC}
\ee
Plugging this into the rest of equations, we find that all
of them are satisfied. Thus, the warping function given in
Eq.~(\ref{WarpsolUC}) is indeed a solution.
Interestingly, this is true for any function of $\tilde{N}(r)$.
Therefore, the conformal lapse function $\tilde{N}(r)$ is unconstrained.
(A similar situation arises also in the spherically symmetric system
investigated in Ref.~\cite{Lu:2009em}.
This particular feature arises due to the specific choice of
coefficients to satisfy the detailed-balance condition.)

For $\Lambda_W <0$,
depending on the signature of $\alpha$, the warp factor in
Eq.~\eqref{WarpsolUC} can be rewritten as follows;
\begin{equation}
W(z)=
\left\{
  \begin{array}{ll}
    \sqrt{\frac{-\Lambda_W}{\alpha}} \sinh \sqrt{\alpha} (z-z_0)
& \hbox{\qquad {\rm for} \quad $\alpha > 0$,} \\
    \sqrt{-\Lambda_W} (z -z_0)
& \hbox{\qquad {\rm for} \quad $\alpha = 0$,} \label{W2}\\
    \sqrt{\frac{\Lambda_W}{\alpha}} \sin \sqrt{-\alpha} (z -z_0)
& \hbox{\qquad {\rm for} \quad $\alpha < 0$.} 
  \end{array}
\right.
\end{equation}
The second solution in Eq.~\eqref{W2} can also be seen by taking the
limit of the third one. For $\Lambda_W >0$ , we have the solution
only when $\alpha >0$ \be W(z) = \sqrt{\frac{\Lambda_W}{\alpha}}
\cosh \sqrt{\alpha} (z -z_0). \label{W4} \ee

By comparing these results with the cases in Einstein gravity in
Eqs.~(\ref{warp2})-(\ref{warp4}), one may identify parameters
as~\footnote{The second relation has a factor 2 difference
from that in Eq.~\eqref{c}, which is similar to the spherical case
in Ref.~\cite{Lu:2009em}.}
\be
\alpha \rightarrow \Lambda_3, \qquad {\rm and} \qquad \Lambda_W
\rightarrow \Lambda_4/3 .
\ee
Therefore, the warp factors seem to be same both in the Einstein and
the Ho\v{r}ava gravity theories.
As it was mentioned above, however, the conformal
lapse function is unconstrained in Ho\v{r}ava gravity.

Note that Eq.~(\ref{Weq}) has another type of solutions, namely,
\be
W(z) = \sqrt{\Lambda_W/\alpha}.
\ee
However, it turns out that there is no $\tilde{N}(r)$ for which all
the rest of equations are satisfied.~\footnote{For the value of
$\lambda =1$, however, we have a constant warp-factor solution as
shall be shown below.} Note also that, in the case of
$\Lambda_W=0$,~\footnote{In this limiting case, Ho\v{r}ava gravity
does not have the Einstein-Hilbert piece becoming a pure
higher-order gravity theory.}
the solution~(\ref{WarpsolUC}) can be reexpressed as
\be
W(z) = \frac{1}{\sqrt{\alpha}} e^{\pm \sqrt{\alpha} (z -z_0)} .
\label{zeroCo}
\ee

Now let us consider the case that Eq.~(\ref{Neq}) is satisfied. The
solution for this equation is in general given by
 \be
\tilde{N}(r) = n_1 \sqrt{-M -\alpha r^2} +n_2,
 \ee
where $n_1$ and $n_2$ are integration constants. With this conformal
lapse function, however, it turns out that there exists no $W(z)$
for which all remaining equations are satisfied, other than the
solutions given in Eq.~(\ref{WarpsolUC}).~\footnote{If $\lambda =1$
is allowed, there exists a solution of constant warping factor with
$n_2 = 0$ though.}
Therefore, we conclude that the warp factor $W(z)$ given by Eq.~\eqref{WarpsolUC}
with the conformal lapse function $\tilde{N}(r)$ unconstrained
is the solution to the field equations.

\subsection{BTZ black-string solution}
In this subsection, we consider the special cases of
$\lambda = 1/3$, $1/2$, and $1$, which were excluded in the previous
subsection. For $\lambda = 1/3$, the theory itself is not defined
well. Thus we do not consider this case. For $\lambda =1/2$, it
turns out that the solution is again
the warp factor exactly given in
Eq.~(\ref{WarpsolUC}) with the unconstrained conformal lapse
function.

For the case of $\lambda =1$, there exist two classes of solutions.
The first class is the same as
that in the previous section; the warp factor is given by
Eq.~(\ref{WarpsolUC}) with the unconstrained conformal lapse
function. Therefore, this class of solutions exists for the
Ho\v{r}ava-Lifshitz gravity
theory with any value of $\lambda$. In fact, the theory parameter
$\lambda$ does not appear in the metric functions at all.

The second class of solutions is given by
\be
W(z) = \sqrt{\frac{\Lambda_W}{\alpha}},
\qquad {\rm and} \qquad \tilde{N}^2(r) = -M
-\alpha r^2.
\label{ConstantWarp}
\ee
Here, an integration constant was absorbed by rescaling the
$t$-coordinate. The conformal lapse function $\tilde{N}(r)$ is not
unconstrained, but is determinative in this case. Note that $\alpha$
must be negative (positive) if $\Lambda_W$ is negative (positive).
In other words, when the four dimensional cosmological constant
$\Lambda_W$ is negative, an effective three dimensional cosmological
constant of positive value $\alpha$ is not allowed.
This property
differs from that of the previous solutions given in Eq.~(\ref{WarpsolUC})
in which both vanishing and positive effective three dimensional
cosmological constants were allowed.

In order to see how the value of $\lambda =1$ is picked out, one may
assume that the warping factor is constant, i.e., $W(z)=W_c$.
The solution to Eq.~\eqref{Ett} is given by
\be
W^2_c = \frac{-1 \pm \sqrt{2(3\lambda -1)}}{2\lambda -1}
\frac{\Lambda_W}{\alpha}.
\label{Wc}
\ee
With this value of $W_c$ the equation $E_{rr}=0$ for the conformal
lapse function can easily be solved, giving
\be
\tilde{N}^2(r) = -M -\alpha r^2.
\label{N}
\ee
The metric functions $f(r)$, $\tilde{N}(r)$,
and, $W_c$ obtained above are the solutions to the remaining
components of the field equation $E_{ij}=0$, while
the left-hand side of $E_{zz}=0$ equation does not vanish in general,
but it is proportional to $(\lambda -1)$ for the case of
plus sign in Eq.~(\ref{Wc}).
Therefore, the solution exists only for $\lambda =1$, and it
becomes,
\be
ds^2 = \frac{\alpha}{\Lambda_W} \left[ -\left( -M -\alpha r^2
\right) dt^2 + \frac{dr^2}{-M -\alpha r^2} +r^2d\theta^2 +dz^2
\right].
\label{BTZstring}
\ee
After rescaling the coordinates,
$(T,R,Z) \equiv \sqrt{\alpha/\Lambda_W} (t,r,z)$,
the solution becomes
\be
ds^2 = -\left(
-M - \Lambda_W R^2\right)dT^2 + \frac{dR^2}{-M -\Lambda_W R^2}
+R^2d\theta^2 +dZ^2.
\label{BTZstring2}
\ee
Note that the cosmological constant is not the four dimensional one
rather than the three dimensional one.~\footnote{This transformation may be regarded as
setting the integration constant $\alpha$ to be $\alpha =\Lambda_W$.}
As it was mentioned below Eq.~\eqref{c},
since $\lambda =1$ now, $\Lambda_W <0$.
(For the model with the analytic continuation, $\Lambda_W>0$.)
If $M>0$, the above solution represents a BTZ black string
which possesses a translational symmetry along the string direction.
If $M<0$, the geometry is static everywhere.~\footnote{For the case
of analytic continuation ($\Lambda_W>0$), the solution with $M>0$
represents a nonstatic geometry everywhere, and the one with $M<0$
represents a symmetrically translated $dS_3$ along the $z$-direction.}

The BTZ black string solution with a translation symmetry is very unique
in Ho\v{r}ava gravity. This type of solution is not possible in Einstein gravity.
The reason why this type of solution is possible can be explained
by considering the effective-matter stress of the higher-order terms in the action.
When there are only Einstein-Hilbert terms,
the spacetime must be isotropic in the presence of a cosmological constant.
The components of the effective-stress tensor $T^i_j$
evaluated from the higher-order terms
$E_{ij}^{(4)}$-$E_{ij}^{(6)}$ in Eqs.~(\ref{E4})-(\ref{E6})
have a relation,
\be
T^r_r = T^\theta_\theta = -\frac{1}{3}T^z_z= -\frac{1}{3}\hat{T}^i_i,
\ee
where $\hat{T}^i_j$ is the effective-stress tensor given solely by the cosmological constant.
The $T^r_r$ and $T^\theta_\theta$ components reduce
the effect of the cosmological constant in the transverse directions,
while $T^z_z$ component adds the effect on the longitudinal direction.
As a result, the components of the total effective-stress tensor,
${\cal T}^i_j = T^i_j + \hat{T}^i_j$,
arrange in such a way
that the translational symmetry is possibly restored,
\be
{\cal T}^r_r = {\cal T}^\theta_\theta = \frac{1}{3}{\cal T}^z_z .
\ee

We can make coordinate transformations further by rescaling
$(\tau,\rho) \equiv (\sqrt{|M|}T,R/\sqrt{|M|})$,
then the metric becomes
\be
ds^2 = -\left( \pm 1 - \Lambda_W \rho^2\right)d\tau^2 +
\frac{d\rho^2}{\pm 1 -\Lambda_W \rho^2} + |M| \rho^2d\theta^2 +dZ^2,
\label{BTZstring2}
\ee
where the upper sign stands for the $M<0$ case.
This metric exhibits the role of the
3D mass-density parameter $M$ (dimensionless);
the transverse geometry is conical.
There exists a deficit angle $\Delta = 2\pi(1-\sqrt{|M|})$ when $|M|<1$.
Note that $M$ is an integration constant of which scale is not limited by the theory.
Therefore, when $|M|>1$, the angle $\Delta$ becomes negative,
which implies a ``surplus angle".~\footnote{A similar situation arises
for the monopole in Ho\v{r}ava gravity investigated in Ref.~\cite{Kim:2009dq}.}


\section{Conclusions}
We searched for string-like solutions in four dimensions based
on Ho\v{r}ava-Lifshitz gravity.
For a restricted class of solutions
where the Cotton-tensor vanishes,
we found that there exist two types of solutions.
The first type of the solutions is warped along the string
string direction, and the conformal lapse function is not constrained.
The well-known warped string-like solutions in Einstein gravity
including the warped BTZ black strings, are the solutions of this type
in Ho\v{r}ava--Lifshitz gravity.
In other words, the solutions in Einstein gravity become the solutions
in Ho\v{r}ava--Lifshitz gravity, but the reverse is not true in general
since the lapse function is not specified.
The parameter $\lambda$ introduced in Ho\v{r}ava--Lifshitz gravity
does not appear in the solution functions at all.
For $\lambda \ne 1$, there exists no other type of solutions
than this one.

The second type of solutions exists additionally only for $\lambda=1$.
In this case, the conformal lapse function is determined.
This solution is uniform along the string direction.
Interestingly, this type of solutions allows a uniform BTZ black string
which is absent in Einstein gravity.
The higher-derivative terms specifically chosen by the detailed-balance
condition in the Ho\v{r}ava--Lifshitz theory
makes this type of solutions possible.
Unlike the spherical case studied in Ref.~\cite{Lu:2009em},
the class of $\lambda$-dependent solutions
does not exist in the axial case.

It is interesting to see if similar properties hold in the
Ho\v{r}ava-Lifshitz gravity theory
in spacetime dimensions higher than four.
The existence of a uniform black string even
in the presence of a bulk cosmological
constant is particularly interesting.
This uniformity is highly nontrivial.
In the four dimensional point of view,
this is a good candidate to test the theory of Ho\v{r}ava-Lifshitz gravity.
In the context of brane world model
it would be very interesting  because the warped
geometry is essential to have the Newtonian gravity on a brane.

\begin{acknowledgments}
Authors are grateful to Hyun-Seok Yang and Mu-In Park for useful
discussions. This work was supported by the Korea Research
Foundation (KRF) grant funded by the Korea government (MEST) No.
2009-0070303 (I.Y.). GK was supported in part by the National
Research Foundation of Korea (KRF) grant funded by the Korea
government (MEST) (No. 2009-0083826) and the Research Grant from SUN
Micro Systems at KISTI. GK would also like to thank the APCTP for
its hospitality.

\end{acknowledgments}

\begin{appendix}*
\section{}
The equations of motion for the action~\eqref{ActionHG} are derived
by variation. The equation from the variation of the lapse function,
$\delta N$, is given by
\be
\frac{2}{\kappa^2}(K_{ij}K^{ij} -\lambda
K^2) + \frac{\kappa^2\mu^2(\Lambda_W R
-3\Lambda_W^2)}{8(1-3\lambda)} +\frac{\kappa^2\mu^2
(1-4\lambda)}{32(1-3\lambda)}R^2 - \frac{\kappa^2}{2w^4} Z_{ij}
Z^{ij}=0,
\label{eom1}
\ee
where
\be Z_{ij}\equiv C_{ij} -
\frac{\mu w^2}{2} R_{ij}.
\ee
The equation from the variation of the
shift function, $\delta N^i$, is given by
\be
\nabla_k(K^{k\ell}-\lambda\,Kg^{k\ell})=0\,.
\label{eom2}
\ee
The equations of motion from the variation $\delta g^{ij}$ are given by
\be
E_{ij} \equiv
\frac{2}{\kappa^2}E_{ij}^{(1)}-\frac{2\lambda}{\kappa^2}E_{ij}^{(2)}
+\frac{\kappa^2\mu^2\Lambda_W}{8(1-3\lambda)}E_{ij}^{(3)}
+\frac{\kappa^2\mu^2(1-4\lambda)}{32(1-3\lambda)}E_{ij}^{(4)}
-\frac{\mu\kappa^2}{4w^2}E_{ij}^{(5)}
-\frac{\kappa^2}{2w^4}E_{ij}^{(6)}=0,
\label{eom3}
\ee
where
\be
E_{ij}^{(1)} &=& N_i \nabla_k K^k{}_j + N_j\nabla_k
K^k{}_i -K^k{}_i \nabla_j N_k-
   K^k{}_j\nabla_i N_k - N^k\nabla_k K_{ij}  \\
&& - 2N K_{ik} K_j{}^k
  -\frac12 N K^{k\ell} K_{k\ell}\, g_{ij} + N K K_{ij} + \dot K_{ij}, \nonumber \\
E_{ij}^{(2)} &=& \frac{1}{2} NK^2 g_{ij}+ N_i \partial_j K+
N_j \partial_i K- N^k (\partial_k K)g_{ij}+  \dot K\, g_{ij},  \\
E_{ij}^{(3)} &=&N(R_{ij}- \frac12Rg_{ij}+\frac32\Lambda_Wg_{ij})-(
\nabla_i\nabla_j-g_{ij}\nabla_k\nabla^k)N,  \label{E3}\\
E_{ij}^{(4)} &=&NR(2R_{ij}-\frac12Rg_{ij})-2 \big(\nabla_i\nabla_j
-g_{ij}\nabla_k\nabla^k\big)(NR),  \label{E4}\\
E_{ij}^{(5)} &=&\nabla_k\big[\nabla_j(N Z^k_{~~i}) +\nabla_i(N
Z^k_{~~j})\big]  -\nabla_k\nabla^k(NZ_{ij})
-\nabla_k\nabla_\ell(NZ^{k\ell})g_{ij},  \label{E5}\\
E_{ij}^{(6)} &=&-\frac{1}{2}NZ_{k\ell}Z^{k\ell}g_{ij}+
2NZ_{ik}Z_j^{~k}-N(Z_{ik}C_j^{~k}+Z_{jk}C_i^{~k})
+NZ_{k\ell}C^{k\ell}g_{ij}  \label{E6}\\
&&-\frac12\nabla_k\big[N\epsilon^{mk\ell}
(Z_{mi}R_{j\ell}+Z_{mj}R_{i\ell})\big] +\frac12R^n{}_\ell\,
\nabla_n\big[N\epsilon^{mk\ell}(Z_{mi}g_{kj}
+Z_{mj}g_{ki})\big] \nonumber\\
&&-\frac12\nabla_n\big[NZ_m^{~n}\epsilon^{mk\ell}
(g_{ki}R_{j\ell}+g_{kj}R_{i\ell})\big]
-\frac12\nabla_n\nabla^n\nabla_k\big[N\epsilon^{mk\ell}
(Z_{mi}g_{j\ell}+Z_{mj}g_{i\ell})\big] \nonumber\\
&&+\frac12\nabla_n\big[\nabla_i\nabla_k(NZ_m^{~n}\epsilon^{mk\ell})
g_{j\ell}+\nabla_j\nabla_k(NZ_m^{~n}\epsilon^{mk\ell})
g_{i\ell}\big] \nonumber\\
&&+\frac12\nabla_\ell\big[\nabla_i\nabla_k(NZ_{mj}
\epsilon^{mk\ell})+\nabla_j\nabla_k(NZ_{mi}
\epsilon^{mk\ell})\big]-\nabla_n\nabla_\ell\nabla_k
(NZ_m^{~n}\epsilon^{mk\ell})g_{ij} .\nonumber
\ee

By plugging the function $f(r)=-M-\alpha r^2$
into the above equations with $N(r,z)=\tilde{N}(r)/W(z)$,
the constraint equation~(\ref{eom1}) associated with lapse
function becomes
\be
\frac{\kappa^2 \mu^2}{8(3\lambda -1)} &&\Big[ \alpha^2 (2\lambda
-1)W^4 -3({\dot{W}}^2 +\Lambda_W)^2 -4\alpha \lambda W^3 \ddot{W}
+4W({\dot{W}}^2
+\Lambda_W)\ddot{W}     \nonumber\\
&&+2W^2 \left[ \alpha ({\dot{W}}^2 +\Lambda_W) +(\lambda
-1){\ddot{W}}^2 \right] \Big] = 0 .
\label{Ett}
\ee
The nonvanishing components in Eq.~(\ref{eom3}) become
\be
&&E_{rr} \times \left( \frac{\kappa^2 \mu^2 \tilde{N}}{16(3\lambda
-1)(-M-\alpha r^2)W^3} \right)^{-1}   \nonumber\\
& & = \frac{(-M-\alpha r^2) \tilde{N}'}{r \tilde{N}} 2W^2 \left[
-\Lambda_W
-\alpha (2\lambda -1)W^2 -{\dot{W}}^2 +2\lambda W \ddot{W}\right]    \nonumber\\
&& -\alpha^2 (2\lambda -1) W^4 -3 \left( \Lambda_W +{\dot{W}}^2
\right)^2 +4W \left( \Lambda_W -(2\lambda -3) {\dot{W}}^2 \right)
\ddot{W}    \nonumber \\
&& +2(\lambda -1) W^2 \left[ \alpha {\dot{W}}^2 +3{\ddot{W}}^2
+3\dot{W} \dddot{W} \right] -2(\lambda -1) W^3 \left( \alpha
\ddot{W} +\ddddot{W} \right) = 0,
\label{Err}
\ee
\be
&&E_{\theta\theta} \times \left( \frac{\kappa^2\mu^2 r^2
\tilde{N}}{16(3\lambda -1)W^3} \right)^{-1}  \nonumber\\
&& =\frac{(-M -\alpha r^2) \tilde{N}'' -\alpha r
\tilde{N}'}{\tilde{N}} 2W^2\left[ -\Lambda_W
-\alpha (2\lambda -1)W^2 -{\dot{W}}^2 +2\lambda W \ddot{W} \right]   \nonumber\\
&& -\alpha^2 (2\lambda -1) W^4 -3 \left( \Lambda_W +{\dot{W}}^2
\right)^2 +4W \left( \Lambda_W -(2\lambda -3) {\dot{W}}^2 \right)
\ddot{W}     \nonumber\\
&& +2(\lambda -1) W^2 \left[ \alpha {\dot{W}}^2 +3{\ddot{W}}^2
+3\dot{W} \dddot{W} \right] -2(\lambda -1) W^3 \left( \alpha
\ddot{W} +\ddddot{W} \right) = 0,
\label{Ethth}
\ee
\be
&&E_{zz} \times \left( \frac{\kappa^2\mu^2 \tilde{N}}{16(3\lambda -1)
W^3} \right)^{-1}  \nonumber\\
&& =- \frac{r(-M -\alpha r^2) \tilde{N}'' +(-M -2\alpha r^2)
\tilde{N}'}{r \tilde{N}} 2W^2 \left[ \Lambda_W -\alpha \lambda W^2
+{\dot{W}}^2 +(\lambda -1)
W \ddot{W} \right]     \nonumber\\
&& +\alpha^2 (2\lambda -1) W^4 -3 \left( \Lambda_W +{\dot{W}}^2
\right)^2 -8 (\lambda -1) W {\dot{W}}^2 \ddot{W}    \nonumber\\
&& +2 W^2 \left[ \alpha \Lambda_W +\alpha (2\lambda -1) {\dot{W}}^2
-(\lambda -1) {\ddot{W}}^2 +2(\lambda -1) \dot{W} \dddot{W} \right]
= 0,
\label{Ezz}
\ee
\be
E_{r\theta} \times \left( \frac{\kappa^2 \mu W}{8w^2 \sqrt{-M
-\alpha r^2}} \right)^{-1}
 =\Big[ r \left( -M -\alpha r^2 \right) \tilde{N}'' +M
\tilde{N}' \Big] \left[ \dddot{W} -\alpha \dot{W} \right] =
0,
\label{Erth}
\ee
\be
E_{rz} \times \left( \frac{\kappa^2\mu^2}{8(3\lambda -1)W}
\right)^{-1} =\left( \lambda -1 \right)\tilde{N}' \left[ -2 \dot{W}
\ddot{W} +W \left( \alpha \dot{W} +\dddot{W} \right) \right] =
0,
\label{Erz}
\ee
\be
&& E_{\theta z} \times \left( -\frac{\kappa^2\mu W\sqrt{-M -\alpha
r^2}}{8w^2r} \right)^{-1} \nonumber\\
&& =\Big[ r^2 \left( -M -\alpha r^2 \right)
\tilde{N}''' +r \left( -M -4\alpha r^2 \right) \tilde{N}'' +M
\tilde{N}' \Big] \left( \ddot{W} -\alpha W \right) = 0,
\label{Ethz}
\ee
By subtracting Eq.~(\ref{Err}) from Eq.~(\ref{Ethth}), one obtains
\be
\Big[ r \left( -M -\alpha r^2 \right)\tilde{N}'' +M \tilde{N}' \Big]
\left[ -\Lambda_W -\alpha \left( 2\lambda -1 \right) W^2
-{\dot{W}}^2 +2\lambda W \ddot{W}\right] = 0.
\label{Eththaa}
\ee
Therefore, one may solve this equation instead of solving
Eq.~(\ref{Ethth}) equivalently.

\end{appendix}

\end{document}